\documentclass [12pt]{article}
\pdfoutput=1

\usepackage{amsmath}
\usepackage{amsfonts}
\usepackage{amscd}
\usepackage{amsthm}
\usepackage{setspace}
\usepackage{amssymb}

\usepackage{graphicx}
\usepackage{authblk}
\usepackage{caption}
\usepackage{ytableau}
\usepackage{mathtools}

\def\Z{\mathbb{Z}}

\def\R{\mathbb{R}}
\def\C{\mathbb{C}}
\def\P{\mathbb{P}}

\def\til{\tilde}

\begin{document}

\begin{titlepage}

\begin{flushright}
KEK-TH-2357
\end{flushright}

\vskip 3.0cm

\begin{center}

{\large Four-dimensional $\mathcal{N}=1$ SCFTs on S-folds with T-branes and\\ AdS/CFT correspondence}

\vskip 1.2cm

Yusuke Kimura$^1$ 
\vskip 0.6cm
{\it $^1$KEK Theory Center, Institute of Particle and Nuclear Studies, KEK, \\ 1-1 Oho, Tsukuba, Ibaraki 305-0801, Japan}
\vskip 0.4cm

\vskip 2cm
\abstract{We study AdS duals of four-dimensional (4D) $\mathcal{N}=1$ superconformal field theories (SCFTs) on $\Z_k$ S-folds with T-branes that flow to 4D $\mathcal{N}=3$ theories. In a previous study, it was discussed how considering T-brane structure breaks the $\mathcal{N}=2$ supersymmetry in 4D SCFTs on $\Z_k$ S-folds down to $\mathcal{N}=1$. In this study, we analyze this effect from the AdS perspective. We also discuss the constructions of 4D $\mathcal{N}=1$ SCFTs on S-folds with T-branes starting from global K3 hypersurfaces whose fibers involve complex multiplications. We choose an approach to directly analyze the global K3 geometry without relying on the standard Weierstrass technique.}  

\end{center}
\end{titlepage}

\tableofcontents
\section{Introduction}
Superconformal field theories (SCFTs) have been extensively studied. SCFTs garner interest because they provide clues for investigating aspects of field theories, including those with strong coupling. 
\par Superconformal symmetry enables the analysis of theoretical structures of SCFTs in detail. This property has been utilized to study four-dimensional (4D) $\mathcal{N}=4$ SCFTs and $\mathcal{N}=2$ \footnote{Pioneering works on 4D SCFTs can be found in \cite{Sohnius1981, Brink1982, Mandelstam1982, Howe1982, Parkes1982, Howe1983, Parkes1983sym, Parkes1983, West1984, Parkes1984, Howe1983uv, Parkes1985, Conlong1993}. The authors in \cite{Howe1989} discussed two-dimensional SCFTs and the Landau-Ginzburg models. In \cite{SeibergWitten199407, SeibergWitten199408}, Seiberg and Witten determined the exact prepotential for low-energy effective 4D $\mathcal{N}=2$ theories. The field has been intensively studied since this progress, e.g., in \cite{Argyres1994, Argyres1995, Argyres199511, Eguchi1996, Minahan199608, Minahan199610, Aharony2007, Argyres2007, Gaiotto2009, Cecotti2010, Argyres2015, Xie2015, Argyres201611, Caorsi2018, Borsten2018, Apruzzi2020, Argyres202003, He202004, Bourget2020, Giacomelli202007, Heckman2020, Giacomelli202010, Martone202102, Cecotti2021}.} SCFTs by using techniques such as superconformal bootstrap and localization. 4D $\mathcal{N}=4$ SCFTs are known to have exactly marginal couplings. 
\par In addition to these classes of 4D SCFTs, 4D $\mathcal{N}=3$ theories were newly constructed in a recent paper \cite{Garcia-Etxebarria2015}. The construction of 4D $\mathcal{N}=3$ theories in \cite{Garcia-Etxebarria2015} used F-theory \cite{Vaf, MV1, MV2} \footnote{There have been various efforts to analyze gauge groups formed in F-theory, e.g., in \cite{MM2014, Morrison2016, Cianci2018, Taylor2019, Kimura201902, Clingher2020, Karozas2020, Angelantonj2020, DelZotto2020, Klaewer2020, Grassi2021, Morrison2021, Raghuram2021}.}, and the authors in \cite{Garcia-Etxebarria2015} considered a finite cyclic group action of $\Z_k$ on the product $\C^3\times T^2$, where $T^2$ denotes a torus. The torus, $T^2$, in the product, $\C^3\times T^2$, must have a specific complex structure so the group action of $\Z_k$ is compatible with the space structure of $\C^3\times T^2$. This geometric compatibility limits the degree $k$ of the cyclic group $\Z_k$ to only finite values, $k=2,3,4,6$. F-theory on the quotient $(\C^3\times T^2)/\Z_k$ probed by D3-branes then yields 4D theories with 12 supercharges \cite{Garcia-Etxebarria2015}. 
\par 4D $\mathcal{N}=3$ SCFTs that do not have $\mathcal{N}=4$ supersymmetry are intrinsically strongly coupled, and do not have marginal parameters \cite{Aharony2015, Cordova2016}. See also, for example, \cite{Nishinaka2016, Aharony2016, Imamura2016, Imamura201606, Agarwal2016, Lemos2016, vanMuiden2017, Bourton2018, Borsten2018} for studies of 4D $\mathcal{N}=3$ theories. 
\par 4D $\mathcal{N}=2$ SCFTs include theories that flow to 4D $\mathcal{N}=3$ theories. Such 4D $\mathcal{N}=2$ SCFTs were constructed in \cite{Apruzzi2020} by placing a probe D3-brane that probes the 7-brane stack in F-theory \footnote{The authors in \cite{Banks1996, Douglas1996, Fayyazuddin1998, Aharony199806, Gukov199808}, for example, discussed D3 branes that probe F-theory singularities.} on $\Z_k$-quotients of an open patch in an elliptically fibered K3 surface times $\C^2$. In contrast to considering $\Z_k$-quotients of the product, flat $\C^3$ times $T^2$, in the construction of 4D $\mathcal{N}=3$ theories in \cite{Garcia-Etxebarria2015}, the operation of replacing $\C \times T^2$ in $\C^3\times T^2$ with an open patch in an elliptic K3 surface (so that the base of the chosen patch is isomorphic to $\C$) to yield 4D $\mathcal{N}=2$ SCFTs on the probe D3-branes in \cite{Apruzzi2020} physically corresponds to the effect of including 7-branes \cite{Apruzzi2020, Heckman2020}. 
\par The 4D SCFTs that flow to 4D $\mathcal{N}=3$ theories are not limited to 4D $\mathcal{N}=2$ SCFTs; indeed, there are also 4D $\mathcal{N}=1$ \footnote{Recent progress of 4D $\mathcal{N}=1$ theories can be found, e.g., in \cite{Bah2011, Gadde2013, McGrane2014, Maruyoshi2016, Borsten2018, Bourton2020, Kimura2020}. Discussions of $\mathcal{N}=1$ S-folds can be found, e.g., in \cite{Borsten2018, Kimura2020}.} SCFTs that also flow to 4D $\mathcal{N}=3$ theories. Such theories were constructed and analyzed in \cite{Kimura2020}. The ``T-brane'' structure \cite{CCHV2010} \footnote{The authors in \cite{DKS2003} studied nilpotent Higgs fields on branes.} was considered in \cite{Kimura2020} to break supersymmetry down to $\mathcal{N}=1$ to construct such theories. T-brane structure involves degrees of freedom on the coincident 7-branes in F-theory, which the Weierstrass equation of an elliptic fibration cannot capture \cite{MT1204}. For recent studies of T-branes, see, for example, \cite{DW2011, MTmatter, MT1204, AHK2013, CR2014, CR2014space, CQV2015, HHRRZ1907, Kimura2020}. 
\par The effect of tilting 7-branes in a stack of parallel 7-branes breaks half of the $\mathcal{N}=2$ supersymmetry down to $\mathcal{N}=1$ as discussed in \cite{Heckman201006, Heckman201009}. The scalar field $\phi$ acquires a vacuum expectation value (vev) that varies over the base space as a result of this operation \cite{Heckman201006}. This effect adds a variation to the superpotential in 4D $\mathcal{N}=2$ SCFTs, which results in breaking $\mathcal{N}=2$ supersymmetry down to $\mathcal{N}=1$ \cite{Heckman201006}. When $\phi$ and $\phi^\dagger$ do not commute, the $\mathcal{N}=2$ supersymmetry is broken down to $\mathcal{N}=1$ as discussed in \cite{Heckman201006, Heckman201009}. In this situation, there is room to consider the T-brane structure \cite{CCHV2010}.
\par We focus on such 4D $\mathcal{N}=1$ SCFTs in this work, which are constructed by breaking half the supersymmetry of 4D $\mathcal{N}=2$ SCFTs on $\Z_k$ S-folds \cite{Apruzzi2020} via T-branes. (Such 4D $\mathcal{N}=1$ SCFTs admit flows to 4D $\mathcal{N}=3$ SCFTs by construction.)

\vspace{5mm}

\par There are two themes in this paper:\\
\begin{itemize}
\item[] 1.  Given SCFTs, it is natural to consider their holographic duals via AdS/CFT correspondence \cite{Maldacena1997, Gubser1998, Witten1998}. The holographic duals of 4D $\mathcal{N}=3$ SCFTs probed by $N$ D3-branes are considered to be F-theory on $AdS_5 \times (S^5\times T^2)/\Z_k$ in the large-$N$ limit \cite{Garcia-Etxebarria2015, Aharony2016}. As mentioned previously, inclusion of 7-branes and considering T-brane structure breaks $\mathcal{N}=3$ supersymmetry down to $\mathcal{N}=1$. We investigate the effect of the T-brane on the AdS side. This is the first theme, and we analyze the holographic duals of 4D $\mathcal{N}=1$ SCFTs. 
\item[] 2.  The other theme is to build global elliptically fibered K3 surfaces to construct $\Z_k$ S-folds. These constructions yield 4D $\mathcal{N}=2$ SCFTs, and considering T-brane structure yields 4D $\mathcal{N}=1$ SCFTs. The constructed 4D $\mathcal{N}=1$ SCFTs are related to 4D $\mathcal{N}=3$ theories. We start with the construction of global K3 geometries, and take finite cyclic group quotients to obtain S-folds. Placing a probe D3-brane probes an infrared (IR) theory, and yields 4D $\mathcal{N}=2$ SCFTs. The effect of T-branes breaks the supersymmetry down to $\mathcal{N}=1$. Because the global aspects of the geometry are considered, this approach enables one to relate the resulting SCFTs to ultraviolet (UV) theories that are in the high-energy regime. We particularly consider hypersurface and complete intersection constructions for global K3 geometries in this study.
\end{itemize}

\vspace{5mm}

\par We mentioned previously that the 4D $\mathcal{N}=1$ SCFTs that we analyze in this study relate to 4D $\mathcal{N}=2$ SCFTs and 4D $\mathcal{N}=3$ SCFTs. Analyzing the holographic duals of such 4D $\mathcal{N}=1$ SCFTs should provide clues for understanding what happens on the AdS side when the $\mathcal{N}=3$ supersymmetry is broken down to $\mathcal{N}=1$. The analysis in this study might help in understanding some aspects of the AdS/CFT correspondence. 
\par F-theory on the product $\C^3 \times T^2$ probed by $N$ D3-branes yields a 4D $\mathcal{N}=4$ SCFT, and the holographic large-$N$ dual of this is the type IIB theory on $AdS_5\times S^5$ \cite{Maldacena1997}. 
\par The construction of 4D $\mathcal{N}=3$ theories in \cite{Garcia-Etxebarria2015} used the $\Z_k$-quotient of the product $\C^3 \times T^2$, and the group action of $\Z_k$ is mathematically consistent only when $k$ takes one of the values 2, 3, 4, or 6. The complex structure of torus $T^2$ in $\C^3 \times T^2$ must also take a specific value to be compatible with $\Z_k$ action \cite{Garcia-Etxebarria2015}. Under these conditions, F-theory on $(\C^3 \times T^2)/\Z_k$ probed by D3-branes yields 4D theory with 12 supercharges \cite{Garcia-Etxebarria2015}. (When $k=2$, F-theory on $(\C^3 \times T^2)/\Z_2$ probed by D3-branes yields a 4D $\mathcal{N}=4$ SCFT.) The holographic dual of the resulting theory is F-theory on $AdS_5\times (S^5\times T^2)/\Z_k$ \cite{Garcia-Etxebarria2015}.
\par When 7-branes are included, the holographic duals of 4D $\mathcal{N}=2$ SCFTs on $\Z_k$ S-folds are also F-theory on $AdS_5\times (S^5\times T^2)/\Z_k$ \cite{Apruzzi2020}. More precisely, the deficit angles of 7-branes need to be included on the AdS side \cite{Apruzzi2020, Heckman2020}. 
\par As one of the themes in this study, we investigate the holographic duals of 4D $\mathcal{N}=1$ SCFTs obtained from 4D $\mathcal{N}=2$ SCFTs on $\Z_k$ S-folds by considering T-brane structure. The effect of considering T-branes corresponds to tilting a stack of parallel 7-branes to a nonparallel configuration. As we will discuss in section \ref{sec3}, under this operation the holographic dual deforms from $AdS_5\times (S^5\times T^2)/\Z_k$ to another geometry. We consider the properties of the resulting dual geometry.  

\vspace{5mm}

\par As to the second theme, we utilize global elliptically fibered K3 surfaces to construct $\Z_k$ S-folds. We then consider T-brane structure to break half the supersymmetry in 4D $\mathcal{N}=2$ SCFTs on the resulting S-folds to yield 4D $\mathcal{N}=1$ SCFTs. We demonstrate that genus-one fibered K3 surfaces built as bidegree (3,2) hypersurfaces in $\P^2\times \P^1$ are useful as a tool in these constructions. Analogous to construction of 4D $\mathcal{N}=3$ theories \cite{Garcia-Etxebarria2015}, one needs to build a genus-one fibered K3 surface whose torus fiber has a suitable complex structure \footnote{Constructions of elliptic K3 surfaces wherein the complex structures of the torus fibers have fixed constant values over the base $\P^1$ were discussed in \cite{Dasgupta1996, Kimura2015, Kimura201603}.} that is compatible with $\Z_k$ action when the $\Z_k$-quotient of the product of an open patch in K3 surface times $\C^2$, $({\rm K3}^{\circ} \times \C^2)/\Z_k$, is considered to yield a $\Z_k$ S-fold. 
\par In fact, use of a mathematical technique realizes a genus-one fibered K3 surface with torus fibers with a suitable complex structure. The technique is to consider genus-one fibered K3 surfaces whose fibers possess {\it complex multiplications}. Constructions of genus-one fibered K3 surfaces whose fibers possess complex multiplications in the context of string theory can be found in \cite{Kimura2015, Kimura201603}. Such K3 surfaces can be useful tools in yielding 4D $\mathcal{N}=2$ and $\mathcal{N}=1$ SCFTs on $\Z_k$ S-folds. A bidegree (3,2) hypersurface in $\P^2\times\P^1$ given by a specific form of defining equation yields the construction of a genus-one fibered K3 surface whose fibers possess complex multiplication of order 3 \cite{Kimura2015}. This construction yields K3 surfaces wherein the complex structures of the torus fibers have fixed constant values over the base $\P^1$, that are compatible with the $\Z_k$ quotient action. We apply this construction of global K3 geometry and analyze 4D $\mathcal{N}=1$ SCFTs on S-folds built from this type of K3 surface. Analogous construction of a K3 surface built as the double covers of $\P^1\times\P^1$ can be found in \cite{Kimura201603}, and 4D $\mathcal{N}=1$ SCFTs on S-folds built from K3 double covers can be found in \cite{Kimura2020}. 
\par Because a probe D3-brane only probes a small neighborhood, the SCFT realized on the probe D3-brane is an IR theory. However, considering the global aspects of K3 geometry, the IR theory of the D3-brane relates to UV theory. Our approach of considering global K3 geometry enables us to relate the IR theories on the probe D3-branes to the behaviors of the theories in the UV limit. Furthermore, the analyses of elliptic K3 surfaces and $\Z_k$ S-folds in this study investigate the geometries directly, without relying on the standard Weierstrass techniques. 

\vspace{5mm}

\par In this work, we focus on the situations where the size of the compactifying space is small, so the resulting theory is approximated by the limit at which gravity decouples. We do not consider the effect of gravity in this note. For 4D theory on the probe D3-brane in the decoupling limit, the symmetry on the 7-branes yields the flavor symmetry in the IR.
 
\vspace{5mm}

\par The remainder of this note is structured as follows: in section \ref{sec2}, we review constructions of 4D $\mathcal{N}=3$ theories \cite{Garcia-Etxebarria2015}, 4D $\mathcal{N}=2$ SCFTs on $\Z_k$ S-folds \cite{Apruzzi2020}, and 4D $\mathcal{N}=1$ SCFTs constructed via considering T-brane structure \cite{Kimura2020}. 
\par In section \ref{sec3}, we investigate the holographic duals of 4D $\mathcal{N}=1$ SCFTs that flow to 4D $\mathcal{N}=3$ theories. This is the first theme of this paper. 
\par In section \ref{sec4}, we study 4D $\mathcal{N}=1$ SCFTs on S-folds that are constructed from K3 hypersurfaces. We start from global geometries of elliptic K3 surfaces built as hypersurfaces in the product of projective spaces. These correspond to the other theme of this study. We state our concluding remarks in section \ref{sec5}.

\section{Reviews of 4D $\mathcal{N}=3$ theories, 4D $\mathcal{N}=2$ and  $\mathcal{N}=1$ SCFTs}
\label{sec2}

\subsection{4D $\mathcal{N}=3$ theories}
\label{subsec2.1}
4D $\mathcal{N}=3$ theories were constructed by the authors in \cite{Garcia-Etxebarria2015}. Here, we briefly review the constructions in \cite{Garcia-Etxebarria2015}.
\par The constructions of 4D theories with 12 supercharges in \cite{Garcia-Etxebarria2015} were generalizations of orientifold 3-planes in type IIB superstring theory. First, we review the orientifold 3-planes. For an orientifold 3-plane in type IIB theory, a point $x$ and its opposite $-x$ in the space transverse to the orientifold 3-plane are identified. Opposite values for NSNS 2-form $B_2$ and RR 2-form $C_2$ are identified in this identification. These identifications act as a $\Z_2$ group on the space transverse to the orientifold 3-plane and the action also involves the element $-I$ in the S-duality group $SL(2,\Z)$. When D3-branes are placed on the orientifold 3-plane, the resulting theory of the D3-branes under the $\Z_2$ action yield a 4D $\mathcal{N}=4$ SCFT. 
\par The story of the orientifold 3-plane generalizes to orbifold constructions, when the $\Z_2$ group is replaced with $\Z_k$, where $k=3,4,6$. Analogous to the orientifold 3-plane, the $\Z_k$ action ($k=3,4,6$) also acts on the space $\C^3=\R^6$ and the $\Z_k$ action involves an element in the S-duality group $SL(2,\Z)$ whose $k$th power yields an identity. Here, it is convenient to consider the F-theory lift, wherein the S-duality group is identified with the automorphism group $SL(2,\Z)$ of the elliptic curve $T^2$ as a fiber. In the lift, the complex structure $\tau$ of the elliptic curve as a fiber is identified with the axio-dilaton $\tau$ in type IIB theory. F-theory on the quotient space under the $\Z_k$ group on the product $\C^3\times T^2$, $(\C^3\times T^2)/\Z_k$, probed by D3-branes yields a 4D $\mathcal{N}=3$ theory \cite{Garcia-Etxebarria2015}. In these constructions, $\Z_k$ action on $\C^3\times T^2$ is mathematically consistent only for $k=2,3,4,6$.
\par The $\Z_k$ group action on $\C^3\times T^2$ is defined as follows \cite{Garcia-Etxebarria2015}:
\begin{equation}
\label{Zk action in 2.1}
(z_1, z_2, z_3, z_4) \rightarrow (e^\frac{2\pi i}{k}\,z_1, e^{-\frac{2\pi i}{k}}\,z_2, e^\frac{2\pi i}{k}\,z_3, e^{-\frac{2\pi i}{k}}\,z_4),
\end{equation}
where $z_1, z_2, z_3$ denote the coordinates of $\C^3$, and $z_4$ denotes the complex coordinate of torus $T^2$ regarded as a complex elliptic curve. 
\par For $k=3,4,6$, the complex structure, $\tau$, of the elliptic curve, $T^2$, namely the axio-dilaton, must take a special value to be consistent with the $\Z_k$ action. For $k=2$, specifically for the case of the orientifold 3-plane, the complex structure $\tau$ can take any value. The values of the complex structures $\tau$ of the elliptic curves $T^2$ to be consistent with the $\Z_k$ action were deduced in \cite{Garcia-Etxebarria2015}; we present the result in \cite{Garcia-Etxebarria2015} in Table \ref{table group and complex structure in 2.1}. 

\begingroup
\renewcommand{\arraystretch}{1.1}
\begin{table}[htb]
\begin{center}
  \begin{tabular}{|c|c|} \hline
Degree $k$ of group $\Z_k$ & allowed complex structure $\tau$ \\ \hline
2 & any value. \\
3 & $e^\frac{\pi i}{3}$ \\
4 & $i$ \\
6 & $e^\frac{\pi i}{3}$ \\ \hline
\end{tabular}
\caption{Allowed complex structure $\tau$ for degree $k$ of finite cyclic group $\Z_k$ in $(T^2 \times \C^3)/\Z_k$ \cite{Garcia-Etxebarria2015} are shown.}
\label{table group and complex structure in 2.1}
\end{center}
\end{table}  
\endgroup

\subsection{4D $\mathcal{N}=2$ SCFTs on S-folds}
\label{subsec2.2}
Utilizing 4D $\mathcal{N}=3$ theories, the authors in \cite{Apruzzi2020} constructed 4D $\mathcal{N}=2$ SCFTs on $\Z_k$ S-folds that flow to 4D $\mathcal{N}=3$ theories. We briefly review the constructions of 4D $\mathcal{N}=2$ SCFTs in \cite{Apruzzi2020}. 
\par The construction in \cite{Apruzzi2020} considered including 7-branes. Including 7-branes changes the geometry of $(T^2 \times \C^3)/\Z_k$ in 4D $\mathcal{N}=3$ theory. Including 7-branes amounts to replacing $T^2\times \C$ in $T^2 \times \C^3$ with an open patch in an elliptically fibered K3 surface, wherein the base space of the open patch is isomorphic to $\C$. The open patch in an elliptic K3 surface is denoted as ${\rm K3}^{\circ}$. Inclusion of 7-branes geometrically distorts the space $(T^2 \times \C^3)/\Z_k$ into $({\rm K3}^{\circ} \times \C^2)/\Z_k$. We use $z_1$ to denote the coordinate of the base curve $\C$ of the open patch ${\rm K3}^\circ$, and we use $z_2, z_3$ to denote the coordinates of $\C^2$ in the product ${\rm K3}^{\circ} \times \C^2$. Therefore, $z_1$ is a coordinate that is transverse to the 7-brane stack, and the coordinates $z_2$ and $z_3$ are parallel to the 7-brane stack.    
\par Placing a probe D3-brane at the $\Z_k$-quotient of a stack of 7-branes in F-theory on $({\rm K3}^\circ \times \C^2)/\Z_k$ yields a 4D $\mathcal{N}=2$ SCFT at low energies. The coordinate $z_1$ transverse to the stack of 7-branes is identified with a chiral superfield $Z_1$ that parameterizes the Coulomb branch (CB). The coordinates $z_2$ and $z_3$, parallel to the 7-brane stack, are identified with decoupled hypermultiplets $Z_2$ and $Z_3$ in 4D SCFT on the D3-brane probe. Including 7-branes, the $\Z_k$-action on $\C^3\times T^2$ (\ref{Zk action in 2.1}) is replaced with the $\Z_k$-action on ${\rm K3}^\circ \times \C^2$ as follows \cite{Apruzzi2020}:
\begin{equation}
(z_1, z_2, z_3) \rightarrow (e^\frac{2\pi i}{k}\,z_1, e^{\phi_2 i}\,z_2, e^{\phi_3 i}\,z_3),
\end{equation}
where the phases $\phi_2$ and $\phi_3$ must satisfy the relation:
\begin{equation}
\phi_2 + \phi_3 = 0 \hspace{0.8cm} {\rm mod} \, 2\pi n
\end{equation}
to preserve $\mathcal{N}=2$ supersymmetry.
\par When a $\Z_k$-action acts on the product ${\rm K3}^\circ\times \C^2$ to build a $\Z_k$ S-fold, analogous to the situations for the constructions of 4D $\mathcal{N}=3$ theories \cite{Garcia-Etxebarria2015}, compatibility conditions are imposed on the modular parameter $\tau$ of the elliptic fibers of the K3 surface and the degree $k$ of the group $\Z_k$. The complex structure $\tau$ of the elliptic fibers of the K3 surface must take a constant value over the base; specifically, the axio-dilaton must be constant over the base, for reasons similar to those for 4D $\mathcal{N}=3$ theories. $\Delta_7$ is used to denote the scaling dimension of the CB operator of a 7-brane type. Because the $\Z_k$ group acts on the base space of the elliptic fibration that is identified with the CB of 4D SCFT on the D3-brane probe, the SCFT on the $\Z_k$ S-fold must have the CB operator of scaling dimension $k\Delta_7$ \cite{Apruzzi2020}. Because the CB operator of the $\Z_k$ S-fold theory yields a phase rotation of supercharges, the authors in \cite{Apruzzi2020} deduced that to preserve supersymmetry, the scaling dimension must take the following values: $k\Delta_7=2,3,4,6$. The $\Z_k$ S-folds constructed by $\Z_k$ groups acting on ${\rm K3^\circ\times \C^2}$ with possible 7-brane types are determined in \cite{Apruzzi2020} as follows \footnote{We used $H_a$, $a=1,2$, to denote the Argyres--Douglas theories \cite{Argyres1995, Argyres199511, Eguchi1996} arising from an SU(2) theory with $a+1$ flavors.}:
\begin{itemize}
\item[] $\Z_2$ S-folds with 7-brane types $E_6$, $D_4$, $H_2$; 
\item[] $\Z_3$ S-folds with 7-brane types $D_4$, $H_1$; 
\item[] $\Z_4$ S-folds with 7-brane type $H_2$. 
\end{itemize}
\par The deformation parameters of the Weierstrass equation of an elliptic fibration are constrained by the geometry of the S-fold quotient \cite{Apruzzi2020, Heckman2020}. Using analysis of $\Z_k$ actions on the Weierstrass equations, the authors in \cite{Apruzzi2020} determined the types of singular fibers over the $\Z_k$-quotients of 7-branes. The mass deformation parameters constitute the Casimir invariants of the flavor symmetry of the $\Z_k$ S-fold theory. Discussions of the flavor symmetries of 4D $\mathcal{N}=2$ SCFTs on S-folds can be found in \cite{Apruzzi2020, Giacomelli202007, Heckman2020}. The rank-one 4D $\mathcal{N}=2$ SCFTs on $\Z_k$ S-folds, $k=2,3,4$, with discrete torsion and the flavor symmetries on the probe D3-branes obtained in \cite{Apruzzi2020} are listed in Table \ref{table rank one SCFTs in 2.2} \footnote{The authors in \cite{Heckman2020} studied rank-one 4D $\mathcal{N}=2$ SCFTs on S-folds without discrete torsion and flavor symmetries.}. 

\begingroup
\renewcommand{\arraystretch}{1.1}
\begin{table}[htb]
\centering
  \begin{tabular}{|c|c|} \hline
S-fold $\Z_k$ quotient of Kodaira fiber & Rank-one 4D $\mathcal{N}=2$ SCFTs \\ \hline
$IV^*$, $\Z_2$ & $[II^*, \, C_5]$ \\ 
$I^*_0$, $\Z_3$ & $[II^*, \, A_3\rtimes \Z_2]$ \\ 
$I^*_0$, $\Z_2$ & $[III^*, \, C_3C_1]$ \\
$IV$, $\Z_4$ & $[II^*, \, A_2 \rtimes \Z_2]$ \\
$IV$, $\Z_2$ & $[IV^*, \, C_2U_1]$ \\
$III$, $\Z_3$ & $[III^*, \, A_1U_1 \rtimes \Z_2]$ \\ \hline
\end{tabular}
\caption{\label{table rank one SCFTs in 2.2}Rank-one 4D $\mathcal{N}=2$ SCFTs on S-folds with discrete torsion by taking $\Z_k$ quotients of singular fibers \cite{Apruzzi2020}. In the left column, items display $\Z_k$ S-fold quotients of singular fibers. $IV, \Z_4$, for example, represents the $\Z_4$ quotient of type $IV$ fiber. In the right column, the first entries in the square brackets are the singular fiber types that correspond to the CB singularities. The groups that appear as the second entries in the square brackets stand for the flavor symmetry groups of the D3-brane probes in the rank-one 4D $\mathcal{N}=2$ SCFTs on $\Z_k$ S-folds. We followed the convention of notation used in \cite{Argyres201611}.}
\end{table}
\endgroup

We list the Kodaira types \cite{Kod1, Kod2} \footnote{The authors in \cite{Ner, Tate} discussed methods to deduce the singular fiber types of an elliptic fibration.} of 7-branes in F-theory in Table \ref{table Kodaira fiber type in 2.2} for the reader's convenience. 

\begingroup
\renewcommand{\arraystretch}{1.1}
\begin{table}[htb]
  \begin{tabular}{|c|c|c|c|} \hline
brane type & axiodilaton (complex structure) & $\Delta_7$ & singularity type \\ \hline
$I_0^*$ & any value. & 2 & $D_4$\\ 
$II^*$ & $e^\frac{\pi i}{3}$ & 6 & $E_8$\\
$III^*$ & $i$ & 4 & $E_7$\\
$IV^*$ & $e^\frac{\pi i}{3}$ & 3 & $E_6$\\
$IV$ & $e^\frac{\pi i}{3}$ & $\frac{3}{2}$ & $A_2$\\
$III$ & $i$ & $\frac{4}{3}$ & $A_1$\\ 
$II$ & $e^\frac{\pi i}{3}$ & $\frac{6}{5}$ & none.\\ \hline
\end{tabular}
\caption{\label{table Kodaira fiber type in 2.2}Kodaira fiber types \cite{Kod1, Kod2} and well-known brane interpretations are presented. $\Delta_7$ denotes the scaling dimension of the CB operator, and $n_7=12\, \frac{\Delta_7-1}{\Delta_7}$ yields the number of 7-branes.}
\end{table}
\endgroup

\subsection{4D $\mathcal{N}=1$ SCFTs on S-folds via T-branes}
\label{subsec2.3}
A class of 4D $\mathcal{N}=1$ SCFTs were obtained in \cite{Kimura2020} by partially breaking the supersymmetry in 4D $\mathcal{N}=2$ SCFTs on S-folds by considering T-brane structure. Let us consider the operation of tilting a parallel 7-brane stack in 4D $\mathcal{N}=2$ SCFTs on S-folds. The tilting operation of the 7-brane stack corresponds to the Higgs field on the 7-branes, $\phi$, gaining a position-dependent, non-zero vev over the base space \cite{Heckman201006}. 
\par S-folds in the construction of 4D $\mathcal{N}=2$ SCFTs \cite{Apruzzi2020} were built as $\Z_k$-quotients of ${\rm K3}^\circ\times \C^2$. As we previously mentioned in section \ref{subsec2.2}, we use $z_1$ to denote the base space $\C$ of the K3 open patch ${\rm K3}^\circ$, and $z_2, z_3$ to denote the coordinates of $\C^2$ in ${\rm K3}^\circ\times \C^2$. 7-branes extend along the directions of $z_2$ and $z_3$. We also stated that, to conform with these notations, we use $Z_2, Z_3$ to represent two $\mathcal{N}=1$ chiral multiplets that parameterize the location of the probe D3-brane that is parallel to the 7-branes. 
\par Tilting a 7-brane stack adds a deformation term, $\delta W=Tr_G (\phi(Z_2, Z_3)\cdot \mathcal{O})$, to the superpotential \cite{Heckman201006}. $\mathcal{O}$ is used in $\delta W$ to represent dimension-two operators in the D3-brane probe theory, which transform in the adjoint representation of the flavor symmetry group \cite{Heckman201006}. As a result, when $\phi$ and $\phi^\dagger$ do not commute, $[\phi, \phi^\dagger]\ne 0$, $\mathcal{N}=2$ supersymmetry is broken down into $\mathcal{N}=1$ \cite{Heckman201006}, yielding 4D $\mathcal{N}=1$ SCFTs as $\mathcal{N}=1$ deformations of 4D $\mathcal{N}=2$ SCFTs on S-folds.
\par When the adjoint-valued vev that the Higgs field $\phi$ acquires does not commute with its conjugate transpose, $\phi^\dagger$, there is room to consider the T-brane structure \cite{CCHV2010}. The condition $[\phi, \phi^\dagger]\ne 0$ is identical to that of the adjoint-valued vev, which the Higgs field $\phi$ takes varying over the base is not diagonalizable by a unitary matrix. A strictly triangular matrix, namely a triangular matrix whose diagonal entries are all zero, yields an example of the vev of the Higgs field $\phi$ that is not diagonalizable by a unitary matrix. 

\subsection{AdS duals of 4D theories}
\label{subsec2.4}
For future use in section \ref{sec3}, we offer a few remarks on the AdS duals of 4D theories. The AdS dual of a 4D $\mathcal{N}=3$ theory probed by $N$ D3-branes in the large-$N$ limit is considered to be type IIB superstring theory on $AdS_5\times S^5/\Z_k$ as discussed in \cite{Garcia-Etxebarria2015}, and the F-theory lift of this is on $AdS_5\times (S^5\times T^2)/\Z_k$ \cite{Aharony2016}. 
\par The holographic duals of 4D $\mathcal{N}=2$ SCFTs constructed by including 7-branes are still considered to be given by F-theory on $AdS_5\times (S^5\times T^2)/\Z_k$ \cite{Apruzzi2020}. To be more precise, deficit angles arising from the effect of including 7-branes also need to be included \cite{Apruzzi2020, Heckman2020}. 
\par What are the holographic duals of 4D $\mathcal{N}=1$ SCFTs obtained as deformations of 4D $\mathcal{N}=2$ SCFTs by considering T-brane structure? Because the supersymmetry is broken down into $\mathcal{N}=1$, it is unpromising to expect that the compact horizon $H$ on the AdS dual would still be given as a $\Z_k$ quotient of the 5-sphere $S^5$. The quotient $S^5/\Z_k$ belongs to a class known as ``Sasaki--Einstein manifolds.'' Because there is $\mathcal{N}=1$ supersymmetry, the deformed horizon $H$, which is due to the effect of the T-brane, stays within the class of Sasaki--Einstein 5-folds \cite{Acharya1998, Morrison1998}.  

\section{Horizon in AdS duals of 4D $\mathcal{N}=1$ SCFTs on S-folds}
\label{sec3}
Here we discuss the holographic duals of 4D $\mathcal{N}=1$ SCFTs on $\Z_k$ S-folds obtained via the T-brane effect from 4D $\mathcal{N}=2$ SCFTs. AdS duals of 4D $\mathcal{N}=3$ theories probed by $N$ D3-branes constructed by using $\Z_k$ orbifolds are type IIB superstring theories on $AdS_5\times S^5/\Z_k$ in the large-$N$ limit \cite{Garcia-Etxebarria2015}. AdS duals of 4D $\mathcal{N}=2$ SCFTs on $\Z_k$ S-folds are type IIB superstring theories on $AdS_5\times S^5/\Z_k$ with 7-branes \cite{Apruzzi2020, Heckman2020}; F-theory uplifts of the dual type IIB theories are theories on $AdS_5\times (S^5\times T^2)/\Z_k$ \cite{Apruzzi2020}. 
\par On the CFT side, the effect of the T-brane structure tilts the parallel 7-brane stacks to a nonparallel configuration, and breaks $\mathcal{N}=2$ supersymmetry down to $\mathcal{N}=1$. What is the effect of this T-brane structure from the perspective of the dual AdS side? We expect that $AdS_5$ space is invariant, and the horizon manifold deforms from the $\Z_k$ quotient of the 5-sphere, $S^5/\Z_k$, to another 5-fold, $H$. We would like to consider the characteristic properties that the horizon $H$ should possess. An immediate property of the horizon $H$ is that it is Sasaki--Einstein 5-fold, because there is $\mathcal{N}=1$ supersymmetry \cite{Morrison1998}. 
\par Before we proceed, we would like to review Sasaki--Einstein manifolds. Given a Riemannian manifold $(M,g)$, one considers the metric cone, $C(M)$, of the Riemannian manifold. Metric cone $C(M)$ is topologically the product of $M$ and the half-line $\R^{>0}$, $M\times \R^{>0}$, and the metric $\til{g}$ on the cone $C(M)$ is given by the following equation:
\begin{equation}
\label{cone metric in 3}
\til{g}=t^2\, g + dt^2,
\end{equation}
where $t$ denotes a coordinate on the half-line $\R^{>0}$. One learns from the metric $\til{g}$ (\ref{cone metric in 3}) that $M$ in the cone $C(M)$ shrinks as $t$ approaches 0. $C(M)$ is literally a ``cone'' whose cross section is the Riemannian manifold $M$. Then $M$ is referred to as Sasaki--Einstein when the metric cone $C(M)$ is K\"ahler and Ricci-flat.
\par The five-dimensional sphere $S^5$ yields a simple example of a Sasaki--Einstein 5-fold. The action of a finite group $G$ on a Sasaki--Einstein manifold $M$ does not alter the Ricci-flat condition, so when $M$ is Sasaki--Einstein, the quotient $M/G$ is also Sasaki--Einstein. This specifically means that $S^5/\Z_k$ is Sasaki--Einstein. 

\vspace{5mm}

\par Now we would like to turn to the discussion of the holographic duals of 4D $\mathcal{N}=1$ SCFTs on $\Z_k$ S-folds. What is the horizon manifold $H$ on the AdS side, when the 7-brane stack is tilted by considering the T-brane structure on the CFT side? 

First, we assume that the $SL_2(\Z)$ bundle is trivial and proceed. It is worthwhile to note that the 4D $\mathcal{N}=1$ SCFTs that we are considering are on the probe D3-brane. 

\par When there is supersymmetry, the metric cone $C(H)$ of the horizon manifold $H$ in the AdS dual of the 4D CFT on the D3-brane has the holonomy group whose connected component has two possibilities: either $SU(3)$ or trivial \cite{Morrison1998} \footnote{The authors in \cite{Morrison1998} deduced these choices from the classification result in \cite{Berger1955}.}. To conform with the notation used in \cite{Morrison1998}, we use $hol^0(C(H))$ to denote the connected component of the holonomy group of the metric cone of the horizon. 
\par We would like to note that given a finite group $G$, the cone of the quotient $S^5/G$ has the holonomy group that is isomorphic to $G$, and this is almost immediate from the definition of the holonomy group. Therefore, when $G$ is finite, the connected component $hol^0(C(S^5/G))$ is trivial. This particularly means that the connected component of the holonomy group $hol^0(C(S^5/\Z_k))$ of the cone of the quotient $S^5/\Z_k$ is trivial. 
\par Under the T-brane effect of tilting the 7-brane stack in 4D $\mathcal{N}=1$ SCFTs on $\Z_k$ S-folds, as the degree to which the 7-brane stack is tilted becomes smaller, the horizon $H$ on the AdS side must approach $S^5/\Z_k$. Because the connected component of the holonomy group $hol^0(C(S^5/\Z_k))$ is trivial, as we stated previously, when the T-brane effect is small, so the geometry of the horizon $H$ is well approximated by $S^5/\Z_k$, the connected component of the holonomy group of the cone of the horizon $H$, $hol^0(C(H))$, should be trivial. Even when the degree to which the 7-brane stack is tilted grows large, the connected component of the holonomy group of the cone of the horizon $H$, $hol^0(C(H))$, suddenly becoming $SU(3)$ is extremely unlikely to occur. At least, the connected component $hol^0(C(H))$ of the holonomy group of the cone of the horizon is unlikely to jump from the trivial group to $SU(3)$ skipping $SU(2)$. For these reasons, one learns that the connected component $hol^0(C(H))$ of the holonomy group of the cone of the horizon $H$ remains the trivial group under the T-brane effect. 

\vspace{5mm}

\par As discussed in \cite{Morrison1998}, when the connected component $hol^0(C(H))$ of the holonomy group of the cone of the horizon $H$ is trivial, the horizon $H$ is isomorphic to a quotient $S^5/\Gamma$, where $\Gamma$ is a finite subgroup of $SU(3)$. When $\Gamma$ is not a subgroup of $SU(2)$, $\Gamma \not\subset SU(2)$, the supersymmetry present in the theory is $\mathcal{N}=1$, while if $\Gamma$ is also a subgroup of $SU(2)$, $\Gamma\subset SU(2)$, the supersymmetry is $\mathcal{N}=2$ \cite{Morrison1998}. 
\par Using this fact, assuming that the $SL_2(\Z)$ bundle is trivial, we deduce that the horizon manifolds in the AdS dual of 4D $\mathcal{N}=1$ SCFTs on $\Z_k$ S-folds are $S^5/\Gamma$, where $\Gamma$ is a finite subgroup of $SU(3)$, $\Gamma\subset SU(3)$, but not contained in $SU(2)$, $\Gamma\not\subset SU(2)$. 
\par The classification of the finite subgroups of $SU(3)$ was discussed by the author in \cite{Blichfeldt}. It is widely known that the finite subgroups of $SU(2)$ are extensions of $ADE$-type groups, because the quotient of $SU(2)$ by $\{\pm 1\}$ is isomorphic to $SO(3)$, $SU(2)/ \{\pm 1\} \cong SO(3)$. $A$-type subgroups correspond to $\Z_k$, and the case where $\Gamma$ is $\Z_k\subset SU(2)$ yields the AdS dual of 4D $\mathcal{N}=2$ SCFTs on a $\Z_k$ S-fold. 

However, the $SL_2(\Z)$ bundle is not trivial for a general S-fold construction. Owing to this situation, determining the horizon requires an additional step for a general S-fold construction.

\section{4D $\mathcal{N}=1$ SCFTs on S-folds and geometry with complex multiplication}
\label{sec4}

\subsection{Main strategy and K3 hypersurfaces with complex multiplication}
\label{subsec4.1}
When S-folds are constructed by utilizing $\Z_k$ group action on a space, one needs to physically consider the $\Z_k$ action on the space and on the axio-dilaton. Consequently, it is convenient to work in the F-theory formulation in which the modular parameter of the torus fiber is identified with the axio-dilaton. Identifying the S-duality group $SL(2,\Z)$ with the automorphism group of the two-torus seen as a complex elliptic curve, the $\Z_k$ action on both the space and the axio-dilaton can be viewed compactly as the $\Z_k$ group action on an elliptic fibration. These two alternative views are identical, but the latter view provides a useful systematic way to analyze physics on S-folds. 
\par F-theory on $\C^3\times T^2$ yields 4D $\mathcal{N}=3$ theories \cite{Garcia-Etxebarria2015}. Inclusion of 7-branes \cite{Apruzzi2020} breaks $\mathcal{N}=3$ supersymmetry down to $\mathcal{N}=2$, and distorts the space. Geometrically, $\C\times T^2$ in $\C^3\times T^2$ needs to be replaced with an open patch in an elliptic K3 surface, ${\rm K3}^\circ$, to represent this situation. F-theory on ${\rm K3}^\circ\times \C^2$ with a probe D3-brane yields 4D $\mathcal{N}=2$ SCFTs \cite{Apruzzi2020}. 
\par For both 4D $\mathcal{N}=3$ and $\mathcal{N}=2$ theories, for $\Z_k$ action defining a mathematically consistent action, the degree $k$ of the group $\Z_k$ must take special values. Except for theories with $\mathcal{N}=4$ supersymmetry, the complex structure of an elliptic fiber must also take a specific value constant over the base space. 

\par For 4D $\mathcal{N}=3$ theories that are not 4D $\mathcal{N}=4$ theories, the axio-dilaton $\tau$ must take the value $exp(\frac{\pi i}{3})$ or $i$, constant over the base space \cite{Garcia-Etxebarria2015}. The situation is analogous for 4D $\mathcal{N}=2$ SCFTs on S-folds with 7-branes included \cite{Apruzzi2020}. 
\par Technically, to construct 4D $\mathcal{N}=2$ SCFTs on $\Z_k$ S-folds, and 4D $\mathcal{N}=1$ SCFTs on S-folds via T-branes, it suffices to construct an elliptic K3 surface whose elliptic fibers have the complex structure $exp(\frac{\pi i}{3})$ or $i$ constant over the base. For the direct product $\C^3\times T^2$, because one can freely choose the complex structure of the torus $T^2$, such construction is obviously possible. However, for elliptically fibered K3 surfaces, the elliptic fiber equations vary over the base complex curve. Consequently, construction of elliptic K3 surfaces with elliptic fibers of either $exp(\frac{\pi i}{3})$ or $i$ complex structure, constant over the base curve, does not seem straightforward. 
\par Genus-one fibered K3 surfaces do not need to be given by the Weierstrass equation, and indeed, there are genus-one fibered K3 surfaces that do not admit such description \footnote{When a genus-one fibered K3 surface does not admit a global section \cite{BM,MTsection}, the surface cannot be described by the Weierstrass equation.}. Only elliptic K3 surfaces that admit a global section can be described by the Weierstrass equations. To maximize the capacity of genus-one fibered K3 surfaces that can be used to construct $\Z_k$ S-folds, and 4D SCFTs on them, it is desirable to develop a method that does not rely on the Weierstrass techniques to build K3 surfaces with elliptic fibers of either $exp(\frac{\pi i}{3})$ or $i$ complex structure that are constant over the base. Fortunately, for application to the construction of $\Z_k$ S-folds, such a method is known in mathematics \footnote{\cite{Kimura201607, Kimura201810, Kimura201902} discussed higher-dimensional generalizations to Calabi-Yau 3-folds and 4-folds whose fibers have constant complex structures $exp(\frac{\pi i}{3})$ and $i$ over the base spaces.}. 
\par Specifically, we consider bidegree (3,2) hypersurfaces in $\P^2\times\P^1$. Such surfaces yield genus-one fibered K3 surfaces; this type of K3 surface was studied in \cite{Kimura2015} in the context of string theory. We particularly focus on the K3 hypersurfaces given by the equation:
\begin{equation}
\label{hypersurface in 4.1}
(t-\alpha_1)(t-\alpha_2)x^3+(t-\alpha_3)(t-\alpha_4)y^3+(t-\alpha_5)(t-\alpha_6)z^3=0.
\end{equation}
The hypersurfaces (\ref{hypersurface in 4.1}) are bidegree (3,2) hypersurfaces in $\P^2\times\P^1$. $t$ denotes the inhomogeneous coordinate on $\P^1$ in $\P^2\times\P^1$, and $[x:y:z]$ yields the homogeneous coordinates of $\P^2$ in $\P^2\times\P^1$. $\alpha_i$, $i=1, \ldots, 6$, are points in $\P^1$. 
\par Equation (\ref{hypersurface in 4.1}) has a highly symmetric form, and owing to this symmetry, the elliptic fibers of the hypersurface (\ref{hypersurface in 4.1}) are the Fermat curve. It is known in mathematics that the Fermat curve has complex multiplication of order 3, and this implies that the elliptic fibers of the hypersurface (\ref{hypersurface in 4.1}), including the singular fibers, have the complex structure $exp(\frac{\pi i}{3})$ \footnote{This is equivalent to the fibers having j-invariant 0.}. 
\par We take an open patch in the K3 hypersurface (\ref{hypersurface in 4.1}) around a point in the base at which a stack of 7-branes is located. As we will discuss in section \ref{subsec4.2}, the K3 hypersurface (\ref{hypersurface in 4.1}) has 7-brane types $H_2$ or $E_6$, depending on the condition on $\alpha_i$. This means that one can construct $\Z_2$ and $\Z_4$ S-folds from the K3 hypersurface (\ref{hypersurface in 4.1}). $\Z_2$ or $\Z_4$ quotients of the chosen open patch, ${\rm K3}^\circ$, times $\C^2$ yields S-folds. Placing a D3-brane probe in F-theory on the resulting S-folds, one obtains 4D $\mathcal{N}=2$ SCFTs on the probe D3-brane. As we will shortly see in section \ref{subsec4.2}, this construction yields 4D $\mathcal{N}=2$ SCFTs $[II^*, C_5]$, $[IV^*, C_2 U_1]$, and $[II^*, A_2\rtimes \Z_2]$ among the SCFTs listed in Table \ref{table rank one SCFTs in 2.2}.

\subsection{4D $\mathcal{N}=1$ SCFTs on S-folds and K3 hypersurface with complex multiplication}
\label{subsec4.2}
To explicitly demonstrate our method, here we use a K3 hypersurface built as a bidegree (3,2) hypersurface in $\P^2\times \P^1$ to construct $\Z_k$ S-folds. Placing a D3-brane probe in F-theory on the resulting S-folds yields 4D SCFTs. For the bidegree (3,2) hypersurface, we focus on the hypersurfaces given by equation (\ref{hypersurface in 4.1}). 
\par The discriminant of the K3 hypersurface (\ref{hypersurface in 4.1}) is given by \cite{Kimura2015}:
\begin{equation}
\Delta \sim \prod_{i=1}^6 (t-\alpha_i)^4.
\end{equation}
In the base $\P^1$ of the K3 surface (\ref{hypersurface in 4.1}), the fibers degenerate to singular fibers at the points $t=\alpha_i$, $i=1, \ldots, 6$. 7-brane stacks are located at these points, and they are wrapped on $\C^2\times \R^{1,3}$. For a generic situation wherein $\alpha$'s are mutually distinct, there are six 7-brane stacks, each consisting of four 7-branes. The singular fibers have type $IV$, and the 7-brane types are $H_2$. 
\par For this generic case, choosing one of the 7-brane stacks and picking an open neighborhood in the K3 surface around the point in the base at which the chosen stack is located, one can construct $\Z_2$ or $\Z_4$ S-folds. Placing a D3-brane probe parallel to the 7-brane stack yields a 4D $\mathcal{N}=2$ SCFT. For the $\Z_2$ S-fold, the flavor symmetry on the 7-branes is $C_2U_1$, and for the $\Z_4$ S-fold, the flavor symmetry is $A_2\rtimes \Z_2$, as previously mentioned in section \ref{subsec4.1}.
\par Considering the T-brane structure partially breaks the supersymmetry, and yields a 4D $\mathcal{N}=1$ SCFT. Because the effect of the T-brane structure corresponds to tilting a stack of 7-branes to make the 7-branes nonparallel, the flavor symmetry algebra of the resulting 4D $\mathcal{N}=1$ SCFT is a subalgebra of the original 4D $\mathcal{N}=2$ SCFT \cite{Kimura2020}. 

\vspace{5mm}

\par When a pair of $\alpha$'s coincide, say $\alpha_1=\alpha_2$, the type of singular fiber over $t=\alpha_1$ becomes $IV^*$, and the 7-brane type is enhanced to $E_6$ \cite{Kimura2015}. If a triplet of $\alpha$'s or more become coincident, which destroys the Calabi--Yau condition. When $\alpha_1=\alpha_2$, picking an open neighborhood of the point $t=\alpha_1$ in the base at which the $E_6$ 7-brane stack is located, one can construct a $\Z_2$ S-fold. The 4D $\mathcal{N}=2$ SCFT on this S-fold is $[II^*, C_5]$. 
\par Analogous to the situations that we discussed previously, considering T-brane structure yields a 4D $\mathcal{N}=1$ SCFT on the constructed S-fold. 

\vspace{5mm}

\par Before we move on to section \ref{subsec4.3}, we would like to offer a few remarks. Because the elliptic fibers of the K3 hypersurface (\ref{hypersurface in 4.1}) possess complex multiplication of order 3, the fibers, including singular fibers, have the complex structure $e^\frac{\pi i}{3}$ \cite{Kimura2015}. This is equivalent to stating that they have j-invariant 0. 
\par The j-invariant of an elliptic fiber uniquely determines its complex structure. The j-invariant is invariant under the isomorphisms of an elliptic curve, and is a function of the modular parameter $\tau$ of an elliptic curve. When an elliptic curve as a fiber is given by the Weierstrass equation $y^2=x^3+f(\tau)x+g(\tau)$, the j-invariant $j(\tau)$ of the curve is expressed using the Weierstrass coefficients as \cite{Cas}:
\begin{equation}
j(\tau)=\frac{1728\cdot 4 \, f(\tau)^3}{4f(\tau)^3+27g(\tau)^2}.
\end{equation}
\par J-invariants of singular fibers of complex elliptic surfaces were determined in \cite{Kod1}. Because the singular fibers of the K3 hypersurface (\ref{hypersurface in 4.1}) have j-invariant 0, there are only five possible singular fiber types: $II$, $IV$, $IV^*$, $II^*$, and $I_0^*$ \cite{Kod1}. This argument specifically excludes two infinite series, $I_n$ $(n\ge 2)$, $I^*_m$ $(m\textgreater 0)$. These are consistent with the actual singular fibers of the K3 hypersurface (\ref{hypersurface in 4.1}) as observed in \cite{Kimura2015}.
\par As another note, the associated Jacobian fibration of the K3 hypersurface (\ref{hypersurface in 4.1}) is given by the following equation \cite{Kimura2015}:
\begin{equation}
y^2=x^3-3^3\cdot 2^4\cdot \prod_{i=1}^6 (t-\alpha_i)^2.
\end{equation}

\subsection{Complete intersection K3 surfaces as a tool to construct S-folds}
\label{subsec4.3}
As we saw, hypersurface constructions of global K3 surfaces can be used to build several examples of S-folds, and 4D $\mathcal{N}=2$ and $\mathcal{N}=1$ SCFTs on the constructed S-folds can be considered. There are multiple other constructions of genus-one fibered K3 surfaces without relying on the Weierstrass techniques, including a construction of K3 surfaces as complete intersections. 
\par For genus-one fibered K3 surfaces built as hypersurfaces and double covers \cite{Kimura2015, Kimura201603}, one can choose members that are useful for construction of S-folds, by selecting the surfaces with elliptic fibers possessing complex multiplications of orders 3 or 4 \footnote{Discussion of genus-one fibered K3 surfaces whose fibers possess complex multiplication of order 4 in the context of string theory can be found in \cite{Kimura201603}.}. We discussed previously that when bidegree (3,2) hypersurface in $\P^2\times\P^1$ is given by the equation of the form (\ref{hypersurface in 4.1}), then its elliptic fibers have complex multiplication of order 3, forcing the fibers to have complex structure $exp(\frac{\pi i}{3})$ constant over the base $\P^1$. This property was useful in constructing $\Z_k$ S-folds from the K3 hypersurfaces. 
\par Here, we consider whether K3 complete intersections also contain a useful class for constructing $\Z_k$ S-folds. This amounts to searching for members in complete intersection K3 surfaces that have a torus fibration, whose elliptic fibers have constant complex structure either $exp(\frac{\pi i}{3})$ or $i$. 
\par There are several ways to build K3 surfaces as complete intersections in projective spaces or in products of projective spaces. To ensure that the resulting K3 surface has a torus fibration, we focus on (2,1) and (2,1) complete intersections in $\P^3\times \P^1$ \footnote{This type of complete intersection K3 surfaces were analyzed in the context of string theory in \cite{Kimura201603}.} \footnote{After a slight appropriate modification, our discussion here also applies to (3,1) and (1,1) complete intersections in $\P^3\times \P^1$.}. This type of K3 surfaces always has a genus-one fibration \cite{Kimura201603}.
\par It appears that there is not an obvious algorithm to select members in K3 surfaces built as complete intersections of two (2,1) hypersurfaces in $\P^3\times \P^1$ with elliptic fibers of constant complex structures $exp(\frac{\pi i}{3})$ or $i$ over the base $\P^1$. However, at least one can show that the family of K3 surfaces built as (2,1) and (2,1) complete intersections in $\P^3\times \P^1$ indeed contains members with elliptic fibers having constant complex structures either $exp(\frac{\pi i}{3})$ or $i$ over the base. We would like to explain this point. 
\par The complete intersection of two (2,1) hypersurfaces in $\P^3\times \P^1$ is given by the following equations:
\begin{eqnarray}
\label{complete intersection in 4.3}
\begin{split}
& b_{1} x_1^2+b_{2} x_2^2+b_3x_3^2+b_4x_4^2+ 2b_5 x_1x_2+ 2b_6 x_1x_3 +2b_7x_1x_4+ 2b_8 x_2x_3 \\
& + 2b_9 x_2x_4 + 2b_{10} x_3x_4 =  0 \\ 
&c_{1} x_1^2+c_{2} x_2^2+c_3x_3^2+c_4x_4^2+ 2c_5 x_1x_2+ 2c_6 x_1x_3 + 2c_7x_1x_4+ 2c_8 x_2x_3 \\
& + 2c_9 x_2x_4 + 2c_{10} x_3x_4= 0,
\end{split}
\end{eqnarray}
where $[x_1:x_2:x_3:x_4]$ denotes the homogeneous coordinates of $\P^3$ in $\P^3\times\P^1$. $b_i$ and $c_j$, $i,j=1, \ldots, 10$, are used to denote homogeneous polynomials of degree 1 on $\P^1$ in $\P^3\times\P^1$.
\par As one may observe, even after reducing a few degrees of redundancies, the complete intersection (\ref{complete intersection in 4.3}) contains many moduli parameters. Owing to this large degrees of freedom, one naturally expects that the whole family of complete intersection K3 surfaces (\ref{complete intersection in 4.3}) would contain members whose elliptic fibers have complex structures either $exp(\frac{\pi i}{3})$ or $i$ constant over the base. We now demonstrate that this expectation is indeed true.
\par The $\tau$ functions and discriminant loci of a K3 genus-one fibration and the Jacobian fibration are identical \cite{MTsection}. Owing to this mathematical fact, it suffices to find members in the moduli of the complete intersections (\ref{complete intersection in 4.3}) whose Jacobian fibrations have elliptic fibers with complex structures $exp(\frac{\pi i}{3})$ or $i$ constant over the base.
\par A method to construct the Jacobian fibration from the complete intersection (\ref{complete intersection in 4.3}) is known. Explanation of this method can be found in \cite{Kimura201603, Kimura201608, Kimura201905, Kimura201908} \footnote{\label{footnote in 4.3}Briefly, one arranges the coefficients, $b_{i}$ and $c_{j}$, of the two hypersurfaces in the complete intersection (\ref{complete intersection in 4.3}) into two symmetric 4$\times$4 matrices, which we denote as $B$ and $C$. One introduces a formal variable $\lambda$, and multiplies one of the two coefficient matrices, $C$, by $\lambda$, then subtracts this matrix $\lambda C$ from the other coefficient matrix, $B$. Taking the determinant of the resulting matrix, $B-\lambda C$, and equating with $u^2$, one obtains a double cover description, $u^2={\rm det}(B-\lambda C)$, of a genus-one fibered K3 surface. The Jacobian fibration of the resulting K3 double cover \cite{BM, MTsection} indeed yields the Jacobian fibration of the complete intersection (\ref{complete intersection in 4.3}) \cite{Kimura201905, Kimura201908}.}. The method provides an efficient formula to compute the Jacobian fibration of the complete intersection, and the method applies to any complete intersection of the form (\ref{complete intersection in 4.3}).  
\par Because the associated Jacobian fibration has a global section, it is given by the Weierstrass equation, $y^2=x^3+f(b_{1}, \ldots, b_{10}, c_{1}, \ldots, c_{10}) x+g(b_{1}, \ldots, b_{10}, c_{1}, \ldots, c_{10})$. The coefficients $f$ and $g$ in the Weierstrass equation of the Jacobian are polynomials in $b_i$ and $c_j$, $i,j=1, \ldots, 10$. 
\par A mathematical argument shows that, the Jacobians belonging to the locus $f=0$ in the complex structure moduli have an elliptic fibration whose elliptic fibers have complex structure $exp(\frac{\pi i}{3})$ constant over the base $\P^1$. Analogously, members belonging to the locus $g=0$ in the complex structure moduli have an elliptic fibration whose elliptic fibers have complex structure $i$ constant over the base. These loci can be expressed as zeros of polynomials in the coefficients $b_{i}$ and $c_{j}$. Therefore, we have demonstrated that the moduli of complete intersection K3 surfaces (\ref{complete intersection in 4.3}) indeed contains the desired members. 
\par Constructions of $\Z_k$ S-folds using appropriate members in the complete intersection K3 surfaces (\ref{complete intersection in 4.3}) and applications to 4D SCFTs are left for future studies. 
\par We remark here that constructing the Jacobian fibration \cite{Kimura201603, Kimura201608, Kimura201905, Kimura201908} of K3 complete intersection (\ref{complete intersection in 4.3}) has also an application to the flavor symmetries in 4D SCFTs. Because the original complete intersection (\ref{complete intersection in 4.3}) and the Jacobian fibration have identical types of the singular fibers, the Kodaira fiber types of the Jacobian fibration yield those of the original K3 complete intersection. The deduced Kodaira fiber types are used to determine the flavor symmetries on the 7-branes in 4D SCFTs on S-folds, that are obtained as $\Z_k$ quotients of an open patch in the K3 complete intersection times $\C^2$.

\section{Concluding remarks}
\label{sec5}
We studied the holographic duals of 4D $\mathcal{N}=1$ SCFTs that are obtained via T-branes as deformations of 4D $\mathcal{N}=2$ SCFTs on $\Z_k$ S-folds. 
\par We also discussed constructions of 4D $\mathcal{N}=1$ SCFTs on $\Z_k$ S-fold, starting form the global geometry of K3 surface. As we studied in section \ref{subsec4.1} and \ref{subsec4.2}, the use of K3 hypersurfaces possessing complex multiplication of order 3 was useful for constructing the 4D $\mathcal{N}=1$ SCFTs. The novelty of our approach is that the method described in this study does not rely on the standard Weierstrass technique, and therefore the method applies to a wider class of K3 surfaces, including those do not admit the Weierstrass description. Furthermore, our method relates constructions of S-folds to global K3 geometries, so CFTs on the probe D3-branes can be related to UV theories that flow to the CFTs in the IR. 
\par Furthermore, we also explicitly demonstrated in section \ref{subsec4.3} that the moduli space of complete intersections contain members that are useful in constructing 4D SCFTs on S-folds.
\par The conformal anomalies $a_{IR}$ and $c_{IR}$ for the 4D $\mathcal{N}=1$ SCFTs constructed in this note can be computed by utilizing the method described in \cite{Heckman201009}, or the method discussed in \cite{Apruzzi201808} with the geometric interpretation \cite{Carta201809}.

\section*{Acknowledgments}

We would like to thank Shigeru Mukai for discussions.

\end{document}